\RequirePackage{fix-cm}
\documentclass[smallextended]{svjour3}       
\smartqed  

\usepackage{color}
\usepackage{graphicx}
\usepackage{revsymb}
\begin{document}

\title{Addressing the cosmological $H_0$ tension by the Heisenberg uncertainty}



\author{Salvatore Capozziello        \and
        Micol Benetti \and
        Alessandro D.A.M Spallicci 
}


\institute{S. Capozziello \at
- Universit\`a degli Studi di Napoli, Federico II (UNINA), Dipartimento di Fisica Ettore Pancini, Via Cinthia 9, 80126 Napoli, Italy, \\ 
- Istituto Nazionale di Fisica Nucleare (INFN), Sezione di Napoli, Via Cinthia 9, 80126 Napoli, Italy, \\
- Gran  Sasso Science Institute (GSSI), Via Francesco Crispi 7, 67100  L'Aquila,  Italy, \\
- Tomsk  State  University  of Control  Systems  and  Radioelectronics (TUSUR), Laboratory  for  Theoretical  Cosmology, 40 prospect Lenina, 634050 Tomsk, Russia.\\
              \email{capozziello@unina.it}           
           \and
           M. Benetti \at
- Universit\`a degli Studi di Napoli, Federico II (UNINA), Dipartimento di Fisica Ettore Pancini, Via Cinthia 9, 80126 Napoli, Italy, \\ 
- Istituto Nazionale di Fisica Nucleare (INFN), Sezione di Napoli, Via Cinthia 9, 80126 Napoli, Italy.\\
              \email{micol.benetti@infn.it}
           \and
           A.D.A.M Spallicci \at
- Universit\'e d'Orl\'eans, Observatoire des Sciences de l'Univers en r\'egion Centre (OSUC) UMS 3116, 1A rue de la F\'{e}rollerie, 45071 Orl\'{e}ans, France, \\
- Universit\'e d'Orl\'eans, Collegium Sciences et Techniques (CoST), P\^ole de Physique, Rue de Chartres, 45100 Orl\'{e}ans, France, \\
- Centre National de la Recherche Scientifique (CNRS), Laboratoire de Physique et Chimie de l'Environnement et de l'Espace (LPC2E) UMR 7328, Campus CNRS, 3A Avenue de la Recherche Scientifique, 45071 Orl\'eans, France, \\
- Universidade do Estado do Rio de Janeiro (UERJ), Instituto de F\'{i}sica, Departamento de F\'{i}sica Te\'{o}rica, Rua S\~ao Francisco Xavier 524, 20550-013 Maracan\~a, Rio de Janeiro, Brazil, \\
- Centro Brasileiro de Pesquisas F\'{\i}sicas (CBPF), Departamento de Astrof\'{\i}sica, Cosmologia e Intera\c{c}\~{o}es Fundamentais (COSMO), Rua Xavier Sigaud 150, 22290-180 Urca, Rio de Janeiro, Brazil.\\
               \email{spallicci@cnrs.orleans.fr}
}

\date{Received: date / Accepted: date}

\maketitle

\begin{abstract}
The uncertainty on  measurements, given by the Heisenberg principle, is a quantum concept usually not taken into account in General Relativity.
From a cosmological point of view, several authors wonder how such a principle can be reconciled with the Big Bang singularity, but, generally, not whether it may affect the reliability of cosmological measurements. 
In this letter, we express the {Compton mass} as a function of the cosmological redshift.  
The cosmological application of the indetermination principle unveils the
differences of the Hubble-Lema\^{i}tre constant value, $H_0$, as measured  from the Cepheids estimates and from the Cosmic Microwave Background radiation constraints. In  conclusion, the $H_0$ tension could be related to the effect of indetermination derived in comparing a kinematic with a dynamic measurement. 
\end{abstract}

\section*{ }
In the last decades, we are experiencing the so-called ``precision cosmology", which stands for the availability of numerous accurate cosmological measurements. Such a 
richness and precision allow confirmations of General Relativity (GR) and of the standard cosmological model, demonstrating that the universe can be our largest available laboratory. At the same time, several tensions and discrepancies emerged, indicating an increasing complexity and the need of adopting huge amounts of unknown and so far undetected dark entities. Observational data point out, substantially, a different evolution (expansion trend) of the universe at different scales.
On the one hand, the increasingly accurate data coming from SNeIa allow a direct estimation of the expansion rate  with high precision, deriving a value of the Hubble-Lema\^{i}tre parameter  $H_0=74.03 \pm 1.42$ \cite{Riess:2019cxk}. 
On the other hand,  detailed Cosmic Microwave Background (CMB) radiation maps, joined 
with  Baryonic Acoustic Oscillations (BAO) data, point to the value $H_0=66.88 \pm 0.92$ \cite{Aghanim:2018eyx}.
This difference, named H$_0$ {\it tension}, of $4.4 \sigma$ \cite{Riess:2019cxk}, is out of any conciliation and may suggest that the two different approaches, namely kinematic and dynamic, are not fully compatible.
In principle, the $H_0$ tension can be  
addressed by examining alternative models to the cosmological standard one, 
or by trying to reduce  data by further treating the systematics more and more accurately. These studies have highlighted additional model troubles, without leading to a definitive solution \cite{Verde:2019ivm,Benetti:2019lxu,Graef:2018fzu,Benetti:2017juy,Bernal:2016gxb}. In conclusion, the debate consists whether we have to invoke new physics or a more detailed data analysis in view of solving the tension.

The discussion is  open and it could lead to revise some fundamental assumptions as discussed, for example, in 
 \cite{Kang:2019azh}. In this letter, we provide another reading of the issue by proposing, in a cosmological context, a discussion of the Heisenberg
indetermination principle, possibly being the straightforward solution of the problem. In other words, the $H_0$ tension could be related to the concept of measurement in a cosmological context.

{In order to start our discussion, it is worth recalling that the Compton wavelength expresses a fundamental limitation on measuring the position of a particle and it is inversely proportional to the particle rest mass, $m_{0}$.
In its reduced form, the Compton wavelength, $\lambdabar_{\rm c}$ is}

%

\begin{equation}
    {\lambdabar_{\rm c}}=\frac {\hbar}{m_{0}c}~,
    \label{eq:Compton}
\end{equation}
where {$\hbar$} is the {reduced} Planck constant and $c$ is the speed of light. 
 It can be interpreted as a fundamental limitation on the measure concept, since it
can be related to the  Heisenberg principle through 

\begin{equation}
\Delta p \Delta x = \Delta E \Delta t = {\Delta (m_0 c^2) \Delta t} > 
{\hbar}/{2}\,,
\label{hp}
\end{equation}
considering the reduced wavelength as a length measurement, $\lambdabar_{c}= \Delta x$, and $m_{0}c = \Delta p$ as a momentum measurement. 
Clearly, these considerations hold for any physical system where a ``length'' and a ``momentum'' can be defined.  

For our purposes, let us stress that a large wavelength  (or a large observation time) corresponds to low-energy photons, which means that the  limit on measurability can be related to the energy (or, inversely, the wavelength) of { the} observed photons. This is intended for a single quantum particle. Instead, for an ensemble of particles, it is possible to independently draw mass upper limits, in presence of a macroscopic {observable}. 

In this perspective, it is useful to recall
 the recent limit on the graviton mass of $ 2 \times 10^{-58}$ kg given by LIGO, when detecting a coalescence of black holes, GW150914, for a signal duration of only 200 ms \cite{abbottetal2016}. 
At its peak luminosity, the black hole binary emitted $10^{80}$ gravitons. If the graviton had an (effective) mass, then the speed of propagation would depend on the wavelength of the radiation.  This would introduce dispersion in the signal, distorting the chirp waveform \cite{will-1998}. The distortion can be bounded using matched filtering. Conversely, a quantum estimate of the same upper limit on the graviton mass would impose an observation lasting more than 32 days. 

\begin{figure*}[!]
\begin{center}
\includegraphics[width=0.45\textwidth]{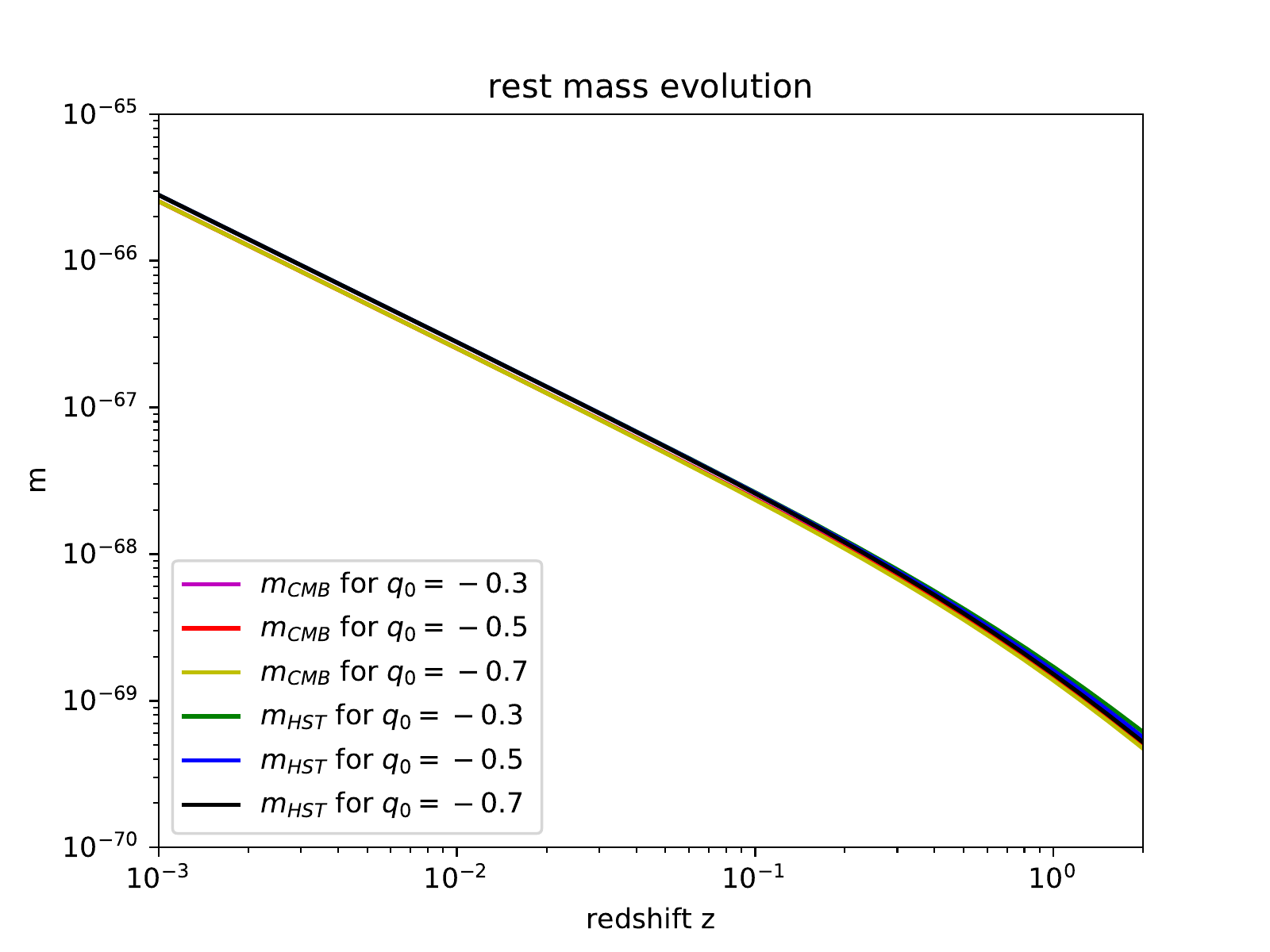}
\includegraphics[width=0.45\textwidth]{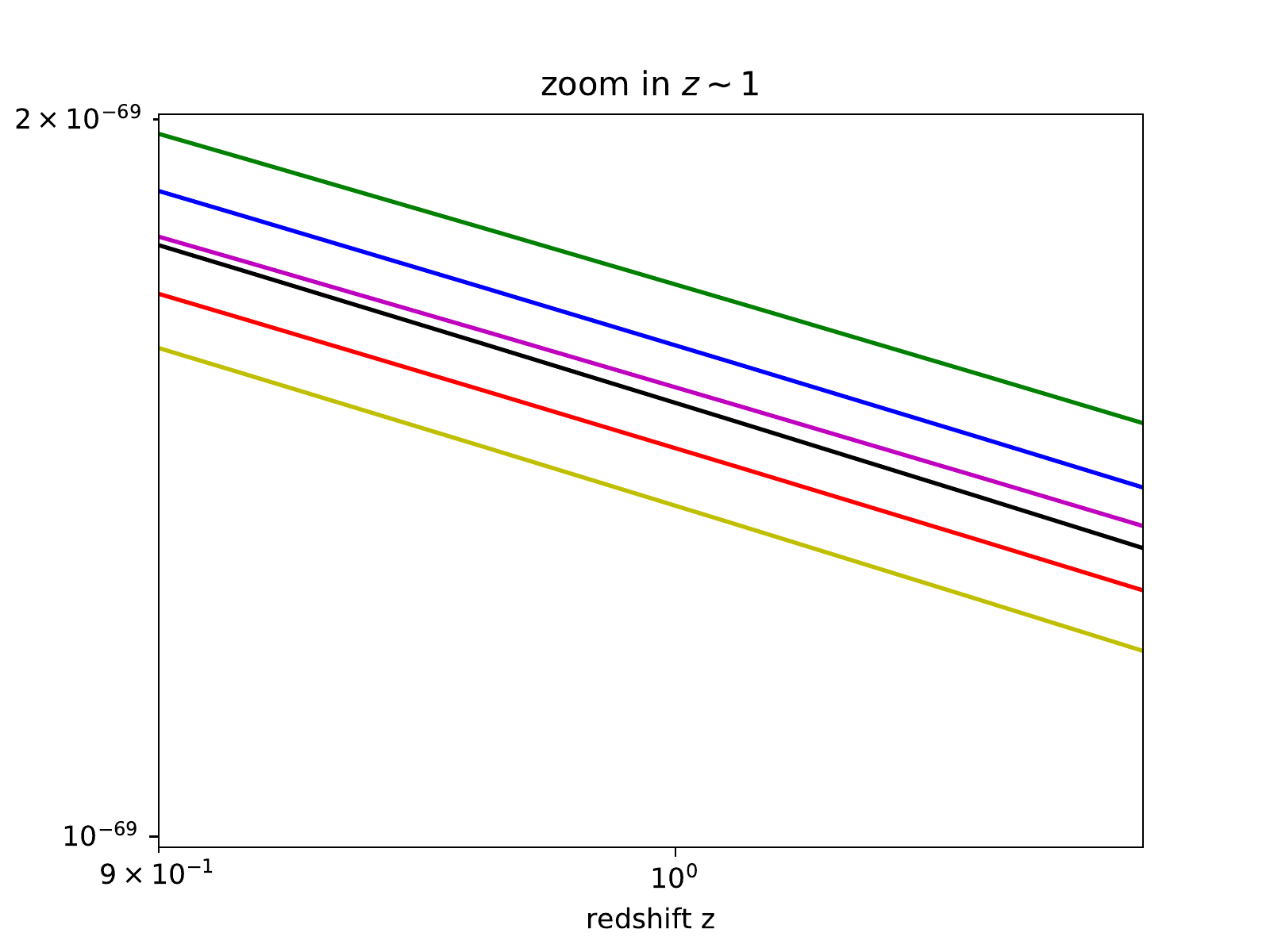}
\caption{Rest mass from Eq. (\ref{eq:m}) as function of the redshift, considering several values of the deceleration parameter and of the Hubble-Lema\^{i}tre constant; $m_{\rm HST}$ refers to $H_0$ = 74 km/s/Mpc, while $m_{\rm CMB}$ to $H_0$=67 km/s/Mpc. 
In the right panel, the {redshift} range $[0.9:1.1]$ is zoomed-in.}
\label{fig:m_z}
\end{center}
\end{figure*}

With these considerations in mind, let us develop  our discussion for homogeneous and isotropic cosmological models  with redshift $z \leq 1$. 
As a leading principle, we will use a cosmographic approach to describe the kinematics of the observed universe. The characteristic Compton length is tantamount to the luminosity length derived from the observations.

Although the above redshift limit can be extended adopting appropriate cosmographic series, for our purposes we use a simple Taylor series. 

It is worth noticing that $ z \sim 1$ is, in some sense, the limit where 
the kinematic description of the pure phenomenological Hubble-Lema\^{i}tre law is superseded by a dynamical description according to the Friedmann-Einstein equations. 
Beyond this value, the simple definition of redshift, coming from the Doppler formula $z \sim v/c $, is no longer valid and the pristine expression  
of the Hubble-Leima\^itre law, $v(z) = H_0 d(z)$, with $v(z)$ the recession velocity of the astrophysical objects at distance $d(z)$, is no longer guaranteed. 
For higher redshifts, we may opt between appropriate (model-independent) cosmographic series as, for example,  considering more detailed  approximants like Pad\'e series \cite{Aviles}, or a dynamical (model-dependent) description of the universe related to cosmological equations of motion. In any case, considerations on  the cosmological  concept of ``measurement'' can be  addressed by simply  adopting a cosmographic Taylor series truncated at the second order where kinematics and dynamics  of Hubble-Lema\^{i}tre flow have to be confronted. 
The luminosity distance $d_L(z)$ is then

\begin{equation}
\label{lux}
d_L(z) = \frac{zc}{H_0} \left[ 1 + \frac{z}{2}(1-q_0) \right]\,,
\end{equation}
where $q_0$ is the cosmographic deceleration parameter of the universe.
This latter is constrained to negative values, {namely either} 
$q_0 = - 0.64 \pm 0.22$ using the joint Pantheon data for SNeIa with BAO and time-delay measurements by H0LiCOW and angular diameter distances measured using water megamasers \cite{Capozziello:2018jya} or $q_0 = - 0.28 \pm 0.49$ using JLA compilation SNeIa data with BAO and observational Hubble{-Lema\^{i}tre} values \cite{Capozziello:2019cav}.

Assuming the luminosity distance as the 
{Compton} wavelength of a particle at redshift $z$, we can substitute Eq. (\ref{lux}) into Eq. (\ref{eq:Compton}) obtaining the corresponding effective ``rest mass'' as

\begin{equation}
m = \frac{\hbar H_0}{zc^2 \left[ 1+ {\displaystyle \frac{z}{2}} (1-q_0)\right] } ~,
\label{eq:m}
\end{equation}
which value of  depends only on  $H_0$ and  $q_0$ and it is decreasing for increasing values of $z$. It is worth stressing that such an effective  mass is derived assuming the Compton length of the observed universe given by the luminosity distance.

For the sake of simplicity in developing our arguments, we choose $z=1$ and arbitrarily fix $q_0=-1/2$. 
Setting the $H_0$ value of the order of the latest measurements of the Hubble Space Telescope (HST) \cite{Riess:2019cxk}, $H_0 \sim 74$ Km/s/Mpc, we find the corresponding mass  $m_{\rm HST}\simeq 1.61 \times 10^{-69}$ kg (we adopt the subscript HST to indicate that this value is obtained by this specific choice of $H_0$). 
Conversely, when we refer to the Hubble-Lema\^{i}tre constant value, derived by 
the Planck measurements of the CMB anisotropies \cite{Aghanim:2018eyx}, for which $H_0 \sim 67$ Km/s/Mpc, we find $m_{\rm CMB}= 1.46 \times 10^{-69}$ kg.

In both cases,  we note an order of magnitude of $10^{-69}$ kg. Such a value is very much lower than the most stringent and official upper limit on the photon mass as reported by the Particle Data Group for a measurement obtained in the Solar System,  $m_{\gamma} < 10^{-54}$ kg \cite{tanabashietal2018}; see \cite{Retino:2013gga} for a critical discussion. According to laboratory tests, the upper bound is instead $m_{\gamma} <1.6 \times 10^{-50}$ kg \cite{Williams:1971ms}. 

A brief reminder on the impact of massive photons on physics is due. Although the first massive theory by de Broglie-Proca of a massive photon is not gauge invariant \cite{debroglie-1936}, later developments have got rid of this feature, considered a limit by  some. Possibly the most popular approach  is the quantizable theory by Stueckelberg \cite{st57} presenting a scalar compensating field.  
Boulware \cite{boulware-1970} showed the renormalizability and unitarity of the Quantum Electro-Dynamics with a de Broglie-Proca massive photon. Supposing that mass rises from the spontaneous symmetry $U(1)$ breaking, gauge invariance is insured also after breaking \cite{itzykson-zuber-2012}, possibly determined by the Higgs mechanism \cite{sgni10}.
Another track was pursued by Guendelman \cite{guedelman-1979} who showed quantization without spontaneous symmetry breaking.
In the Standard-Model Extension, a gauge invariant effective photon mass emerges \cite{bodshnsp2017,bodshnsp2018}. Its mass is proportional to the Lorentz-(Poincar\'e-)Symmetry Violation (LSV). Incidentally, the photon energy-momentum tensor is not conserved in presence of an external Electro-Magnetic (EM) field exchanging energy with the LSV vector or tensor and the EM field.   

Concerning the electric charge conservation, we observe the following. In the de Broglie-Proca modified Gauss law, the coupling of the photon mass to the scalar potential implies a density of ``pseudo-charge'' proportional to the squared mass, added to the ordinary charges. Since the pseudo-charges represent somewhat the Higgs field which does not couple to the ordinary charges, the two kinds of charges are conserved separately. Generally speaking, if the electric flux out of
a surface gives the total electric charge enclosed,
then that charge must be locally conserved: the only way charge can
change is by changing the flux at the same time, and for
a distant surface that flux could not change instantaneously,
if the charge changed. But we cannot impede a charge to change or decay. The mechanism behind charge change or decay should be included in the physical description \cite{nussinov-1987}.

From Eq. (\ref{hp}), referring to the Heisenberg uncertainty in the energy-time form, we can assume  $\Delta t= 13.8$ Gy, the age of the universe, which is the largest possible observation time. Accordingly, we obtain that the smallest measurable mass is $1.35 \times 10^{-69}$ kg, which is slightly below $m_{\rm HST}$ or $m_{\rm CMB}$ but of the same order of magnitude. 

At the same time, we can represent the behaviour of the ``rest mass'' in  Eq. (\ref{eq:m}), 
Fig. (\ref{fig:m_z}), assuming different values of $q_0$. 
If, for larger redshift the ``rest mass'' decreases, looking at low  redshifts, the value increases, nevertheless remaining much below the upper bound on the photon mass previously given, { that is} 
for $z = 0.1$ and still $q_0 = - 1/2$, we get $m_{\rm HST}= 2.61 \times 10^{-68}$ kg and $m_{\rm CMB}= 2.37 \times 10^{-68}$ kg. 

Finally, we have that 
\begin{equation}
\Delta m|_{z=1} = \left. m_{\rm HST}-m_{\rm CMB}\right |_{z=1} \sim 1.5 \times 10^{-70}~{\rm kg}\,.
\end{equation}

It means that a measurement of the photon mass {only beyond} 
the Heisenberg uncertainty can reconcile, in principle, the  $H_0$ tension at $z=1$. Specifically, 
a mass measurement of one order of magnitude { below} 
the indetermination limit 
can solve the current $H_0$ tension at $4.4 \sigma$ \cite{Riess:2019cxk}.
This fact can be formulated also by  writing

\begin{equation}
\frac{\Delta m}{m} =\frac{\Delta H_0}{H_0} \sim 0.1 \,.
\end{equation} 


We can infer that the tension on the $H_0$ measurements can be the effect of the uncertainty of measurement itself at cosmological scales. 
In other words, the uncertainty on the photon mass, derived from a kinematic measurement 
and 
from dynamics, can be the reason of the $H_0$ tension. 

Finally, assuming the photon mass rigorously equal to zero is not correct from the Heisenberg principle point of view  and from the recent analysis of the { Standard-Model Extension} \cite{bodshnsp2017,bodshnsp2018}. According to this consideration, the $H_0$ tension could not require any new physics but only a more accurate discussion of measurements.

\begin{acknowledgements}   
SC and MB acknowledge {the} Istituto Nazionale di Fisica Nucleare (INFN), sezione di Napoli, {\it iniziative specifiche} MOONLIGHT2 and  QGSKY. 
ADAMS acknowledges the Erasmus+ programme for visiting the Universit\`a di Napoli, SC the {Universit\'e d'}Orl\'eans and Campus France for the hospitality. The authors thank the referees and J. A. Helay\"{e}l-Neto for precious suggestions which allowed to improve the paper.

\end{acknowledgements}

%
%



\end{document}